# A Novel 1D Generative Adversarial Network-based Framework for Atrial Fibrillation Detection using Restored Wrist Photoplethysmography Signals

Faizul Rakib Sayem[1,2], Mosabber Uddin Ahmed[1,*], Saadia Binte Alam[3], Sakib Mahmud[2], Md. Mamun Sheikh[1], Abdulrahman Alqahtani[4,5], Md Ahasan Atick Faisal[2], Muhammad E. H. Chowdhury[2,*]

[1]Department of Electrical and Electronic Engineering, University of Dhaka, Dhaka 1000, Bangladesh. Email: faizulrakib-2017214832@eee.du.ac.bd (FRS), mosabber.ahmed@du.ac.bd (MUA), mamunsheikhdueee52@gmail.com (MMS)

[2]Department of Electrical Engineering, Qatar University, Doha 2713, Qatar. Email: mchowdhury@qu.edu.qa (MEHC), sakib.mahmud@qu.edu.qa (SM), atick.faisal@qu.edu.qa (AF)

[3]Department of Computer Science and Engineering Independent University, Bangladesh (IUB), Dhaka 1229, Bangladesh. Email: saadiabinte@iub.edu.bd (SBA)

[4]Department of Biomedical Technology, College of Applied Medical Sciences in Al-Kharj, Prince Sattam Bin Abdulaziz University, Al-Kharj 11942, Saudi Arabia. Email: ama.alqahtani@psau.edu.sa (AA)

[5]Department of Medical Equipment Technology, College of Applied, Medical Science, Majmaah University, Majmaah City 11952, Saudi Arabia

*Corresponding authors: Muhammad E. H. Chowdhury (mchowdhury@qu.edu.qa), Mosabber Uddin Ahmed (mosabber.ahmed@du.ac.bd)

# Abstract

Atrial fibrillation (AF) poses an increased stroke risk, necessitating effective detection methods. While electrocardiogram (ECG) is conventionally used for AF detection, the simplicity and suitability for long-term monitoring make photoplethysmography (PPG) an attractive alternative. In this study, we present a novel approach for AF detection utilizing smartwatch-based wrist PPG signals. Notably, this is the pioneering use of 1D CycleGAN (Generative Adversarial Network) for reconstructing 1D wrist PPG signals, addressing the challenges posed by poor signal quality due to motion artifacts and limitations in acquisition sites. The proposed method underwent validation on a dataset comprising 21,278 10s long wrist PPG segments. Our classification model, Self-AFNet, incorporating 1D-CycleGAN restoration, demonstrated accuracy at 96.41% and 97.09% for the two splits, respectively. The restored signals exhibited a significant accuracy improvement (2.94% and 5.08% for test splits, respectively) compared to unrestored PPG. Additionally, AF detection using ECG signals, paired with matched PPG signals, confirmed the validity of employing reconstructed PPG for classification. Self-AFNet achieved impressive accuracies of 98.07% and 98.97%, mirroring the performance of AF detection using reconstructed PPG segments. This study establishes that reconstructed wrist PPG signals from wearable devices offer a reliable means for AF detection, contributing significantly to stroke risk reduction.

**Keywords:** Atrial Fibrillation; Photoplethysmogram (PPG); Wrist PPG; Electrocardiogram (ECG); Signal Reconstruction; Classification.

## 1. Introduction

Atrial fibrillation (AF) is the most prevalent cardiac arrhythmia globally. As per the 2017 Global Burden of Disease research [1], there were 37.57 million instances of AF globally, accounting for around 0.51% of the global population. AF is distinguished by an irregular heartbeat caused by unregulated electrical impulses from the atria and throughout the heart. The heart muscle can get overworked by excessive or irregular heartbeats, which can result in heart failure (Figure 1). Long-term AF may cause inadequate blood perfusion, leading to fatigue, clots, strokes, and even death.

The chance of developing AF is influenced by age and other genetically predisposed and lifestyle-related diseases and disorders, such as hypertension [2]. Additionally, several studies [3]–[5] indicate that engaging in prolonged physical activity can increase the risk of having an AF episode. AF episodes often become more frequent and more prolonged over time. Neither treatment nor a cure can entirely reverse AF, even though various drugs may help minimize its symptoms. However, owing to the dynamic characteristics of this field of research, rapid and precise AF diagnosis has received increased attention.
Throughout the last several decades, various methods for diagnosing AF have been used and are still in use in medical settings based on electrocardiography (ECG) [7], ballistocardiogram (BCG) [8], and photoplethysmography (PPG) signals [9], [10]. Among them, ECG and PPG signals are most widely used to efficiently detect AF including single-lead ECG, wrist-worn wearable, and standard 12-lead ECG signals [11]. Despite these technologies, long-term patient monitoring techniques can gather more information and enable better detection of transient arrhythmia events, thereby improving the overall detection accuracy of AF. There are several methods and clinical tools, including Holter monitoring system to record ECGs, and clinical PPG signals, are used that may record and track an AF episode. Single-lead and multi-lead [12] ECG signals are studied in the detection of AF with promising accuracy despite being uncomfortable for the patients. Several pre-processing techniques can denoise ECG signals, including Fourier cosine series operation to remove baseline wander and high-frequency components [13]. Noise reduction techniques include the use of a wavelet transform as well as an elliptical band-pass filter [14], [15]. Besides, Z-score normalization and a high-pass filter [16] are often employed for amplitude scaling and offset impact removal, respectively, to standardize the ECG signals for analysis.

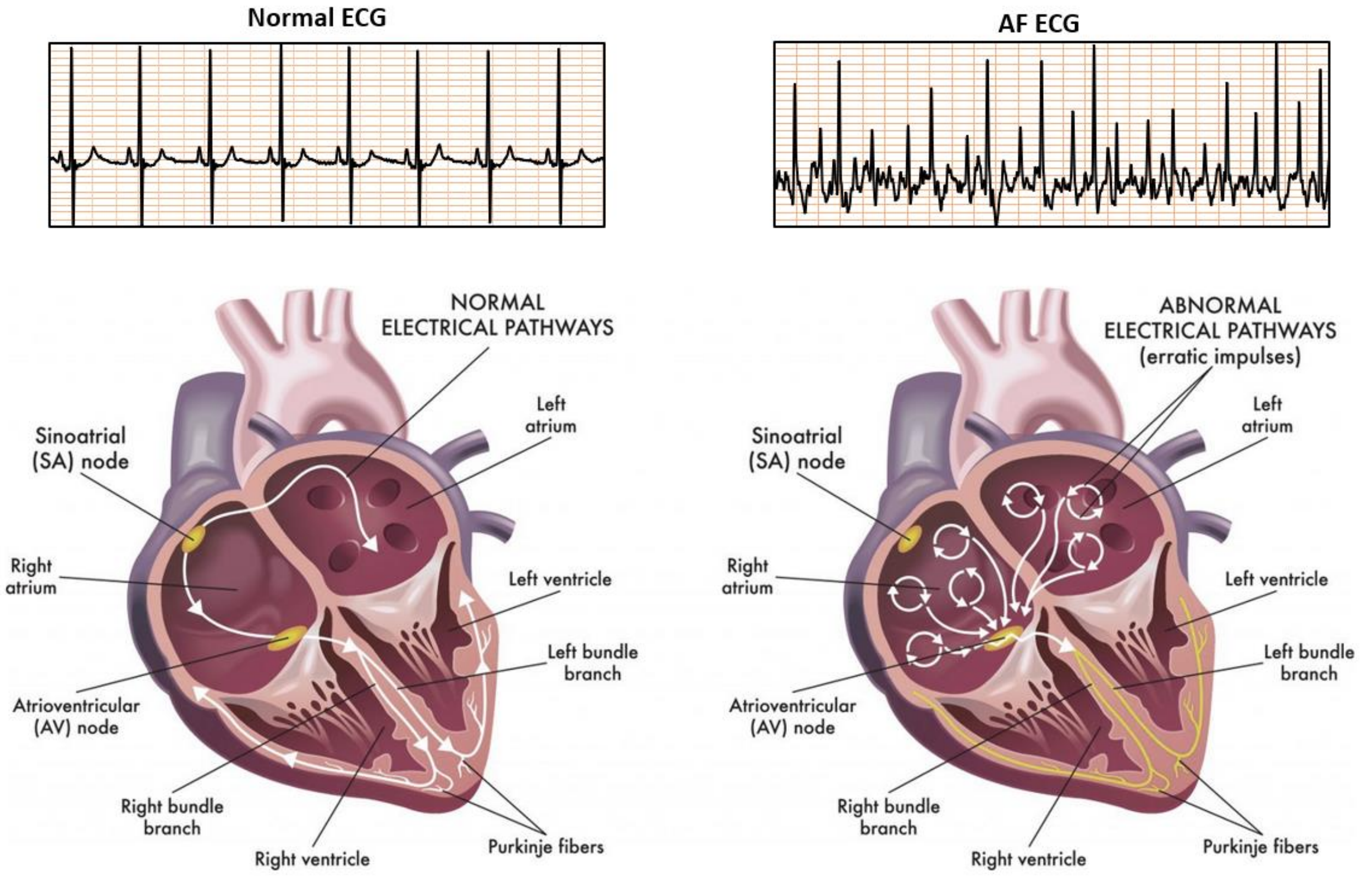

*Figure 1: Illustration of normal sinus rhythm ECG signal with healthy heart (left) and ECG signal with heart suffering from AF (right) (adapted from* [6]*).*

Various robust machine learning based approaches have been used to diagnose arrhythmia from ECG signals. In [17], authors have proposed a multiscale grouped convolutional neural network (CNN) for atrial fibrillation detection and found 97.07% accuracy using 5-fold cross-validation. Zahid et al. [18] proposed a 1D Self-Organized Neural Network (Self-ONN) model and achieved 99.21% accuracy in ECG classification using the MIT-BIH arrhythmia database to distinguish normal, Supraventricular ectopic (S) and Ventricular ectopic (V) beats. In [19], authors used 1D CNN model and transfer learning on both ECG and wrist PPG datasets. They have achieved 95.5% accuracy in AF classification using 5-fold cross-validation method on ECG data and 95.1% accuracy on unseen PPG data. For PPG classification, they have used a very limited number of samples. Several studies show that the detection accuracy for ECG classification may be degraded for many reasons such as screening, sex-specific criteria, and different age ranges [20],[21].

For the aforementioned issues and realistic wearable applications, PPG signal acquired from the wrist has more potential and usefulness for AF detection. Numerous physiological factors might potentially be derived from the analysis of a PPG signal since brain, cardiac, and respiratory interactions regulate blood flow [22]. This has made the resulting PPG signal very rich in carrying physiological information [23]. Therefore, PPG signals are being use in a wide range of applications. A typical AF detection method extracts features from the collected PPG signal (temporal, spectral, or morphological) and analyzes them to determine whether an AF rhythm is present or not. With promising outcomes, PPG time series and their pictorial representation (such as short-term Fourier transform spectrum, or wavelet spectrum) have been applied to detect physiological events [23]–[25]. Bashar et al. [26] proposed motion and noise artifacts (MNA) detection algorithm to extract only clean wrist PPG signals from Umass and Chonlab datasets and performed AF detection on the cleaned data, however, they have used only 366 clean PPG segments (detected by MNA) from both datasets for AF classification. In [27], authors have used optimized signal quality assessment (SQA) technique and obtained promising results (slightly higher than 0.93 accuracy) on the classification task. They have used both wrist PPG and finger PPG from different databases.

It is a very challenging task to acquire PPG signals free of motion artifacts (MA) caused by the subject's deliberate or involuntary movements. Long-term wrist-band PPG recording is not free from such types of noise and motion artifacts, particularly the signal cuts. Even though many signal processing techniques were extensively used to eliminate MA and signal cut conditions, the results were not always satisfactory. As a result, a relatively high percentage of false positives associated with ambulatory PPG signal-based AF detection and a decrease in specificity for the individuals with paroxysmal AF can be easily anticipated.

Despite numerous studies on ECG-based arrhythmia classification, wrist PPG-based AF detection is still mostly undiscovered. Several statistical, machine learning, and deep learning models can be studied to resolve these problems. Generative Adversarial Network (GAN) are outperforming in the field of image synthesis since 2014 [28]. Cycle-Consistent GANs (CycleGANs) [29], [30] are used to translate uncorrelated datasets. 1D CycleGAN for the first time was used in our previous work for blind restoration of ECG signal acquired from Holter monitor [30]. Motivated from that work, in this work, we have implemented 1D CycleGAN to transform corrupted wrist PPG signals into clean PPG signals. Since Cycle-GANs are consistent in preserving the major patterns of the translated signals, all the characteristics of PPG signals are preserved, and we have obtained the improved and denoised PPG signals using the blind restoration. Self-Organized Operational Neural Networks (Self-ONN) [31]–[34] are non-linear neuronal models that can achieve diverse and increased learning capabilities compared to their linear counterpart in conventional CNN model [33]–[35]. There has been no robust study that used a large amount of wrist PPG data to train a classification model for AF detection using wrist PPG signals. Moreover, the low signal quality of the data needs the data to be processed, otherwise, the performance of the classification model will significantly degrade. The novel and most significant contributions of this paper are as follows:

- This is the first study attempting to remove motion artifacts and baseline drifts and restore wrist PPG signals with the aid of **1D CycleGAN** and also the first attempt to use any reconstructed signals for robust AF detection.
- Instead of using conventional CNN layers, we have proposed the recent heterogeneous and nonlinear network model, Self-ONNs to build 1D restoration model**, 1D-CycleGAN,** and 1D classification model, **Self-AFNet**.
- This is the first study to provide several state-of-the-art entropy measurement functions in order to quantitively evaluate the quality of performance of signal restoration of our **1D CycleGAN** model.

- We have proposed a novel **1D Self-AFNet** model for an efficient atrial fibrillation detection using wrist PPG signals andthe performance is compared with the corresponding ground truth ECG signal-based AF detection.

The rest of the paper is organized as follows: Section 2 describes the Dataset Description and Preprocessing. Section 3 shows the Methodology including the theory behind 1D Self-ONNs, Self-AFNet, and 1D Operational CycleGAN. Section 4 contains the experimental setup, experimental result, and performance evaluations of our proposed method on both ECG and PPG signals using various evaluation metrics. Finally, a brief conclusion is presented in Section 5.

## 2. Dataset Description

An open-source dataset called "Open Access Infrastructure for Research in Europe" (OpenAIRE) [36] was used in this study. The dataset consists of long-term ECG and PPG monitoring signals from 8 individuals for 5-8 days with suspected AF rhythms. The dataset contains about 1306 hours of ECG, PPG, and accelerometer data. The ECG and PPG segments were labeled by the expert. It is worth mentioning here that the ECG signal is used for the ground truth annotation of the ECG segments. This publicly accessible dataset was used to classify AF from ECG and PPG signals, which were collected using various sensors. In this study, more than hundred hours of data were used for the training and evaluation of our proposed integrated model. In Table 1, the partition of ECG and PPG recordings and the total number of samples for each class are presented.

**Table 1:** Dataset description for each class for two splits (acceptable wrist PPG)

| | Non-AF samples | AF samples | Total samples |
|---|---|---|---|
| **Test Split 1** | 5846 | 3958 | 9804 |
| **Test Split 2** | 7993 | 3481 | 11474 |
| **Total samples** | 13839 | 7439 | 21278 |

Two experiments were conducted to compare the performance of 1D Self-AFNet model in training one split and testing on another unseen test split. This represents the real-world scenerio and validates the robustness of the proposed model. The PPG samples of Test Split 2 in Table 1 were first considered as unseen test subjects, while Test Split 1 was used for model training. After that, Test Split 2 will be used for training the model whereas Test Split 1 will serve as a test set.

### *2.1. Data Preprocessing*

The ECG and wrist-band PPG data are not suitable for direct use with deep learning models because they are highly susceptible to noise and corruption. To clean the data and get rid of various artifacts (such as baseline drift and motion artifacts), several pre-processing steps are employed before training. ECG and PPG data were first resampled at 250 Hz from 500 Hz and 125 Hz, respectively. Kwon et al. [37] showed that a 250 Hz sampling frequency for ECG signals would provide excellent performance for frequency domain analysis and heart rate variability (HRV) analysis. For the classification and restoration tasks, we employed 10s PPG segments and matching ECG records. The signal length is therefore 2500 samples for each example.

The classification model performs typically poorly if baseline drift is not properly removed. Baseline drift is a low-frequency signal variation problem that results in unintentional amplitude shifts in the signal. Baseline drift correction was carried out utilizing MATLAB's built-in 'movmin' [38], 'polyfit' [39], and 'polyval' [40] functions. Initially, an array of approximated minimum points serving as a baseline approximation for the waveform was found using the moving minimum function. The higher-order polynomial was then fitted using the "polyfit" function together with the estimated points, and the "polyval" function was used to create the polynomial based on the "polyfit" result, which is the estimated baseline. The baseline drift-adjusted signal was then created by subtracting the baseline from the raw signal.
After baseline correction, we performed z-score normalization, followed by a range normalized between 0 and 1 per segment. Band-pass filtering of the ECG signals is very crucial to ensure the band-limited ECG signals are used for further investigation. We have employed a band-pass infinite impulse response (IIR) filter with a cutoff frequency of 0.05 Hz and 100 Hz for removing any potential undesirable noises. For PPG signals, a band-pass filter with cut-off frequencies of 0.5Hz and 25Hz was used.

## 3. Methodology

The proposed AF classification pipeline is illustrated in Figure 2. The wrist PPG and ECG signals were preprocessed. Some of the wrist PPG signal segments were corrupted in such a level that they cannot be used for training CycleGAN for restoration and later on classification. A simple binary classifier was trained using the acceptable and corrupted PPG signals to automatically remove the corrupted signals. The corrupted PPG signals and their accompanying ECG signals segments were eliminated. After this automatic data cleaning process, ECG signals were directly fed into the Self-AFNet model. On the other hand, the acceptable wrist PPG signals were grouped into two groups: good quality and low-quality PPG signals. The low-quality wrist PPG segments were first restored using operational CycleGAN model and then fed into the Self-AFNet model for reliable AF classification. AF detection was performed using ECG and wrist PPG signals were compared independently. The symbol of (1, 3, 1) in the Self-AFNet model refers to the kernel size of the Self-ONN layers described in Figure 5.

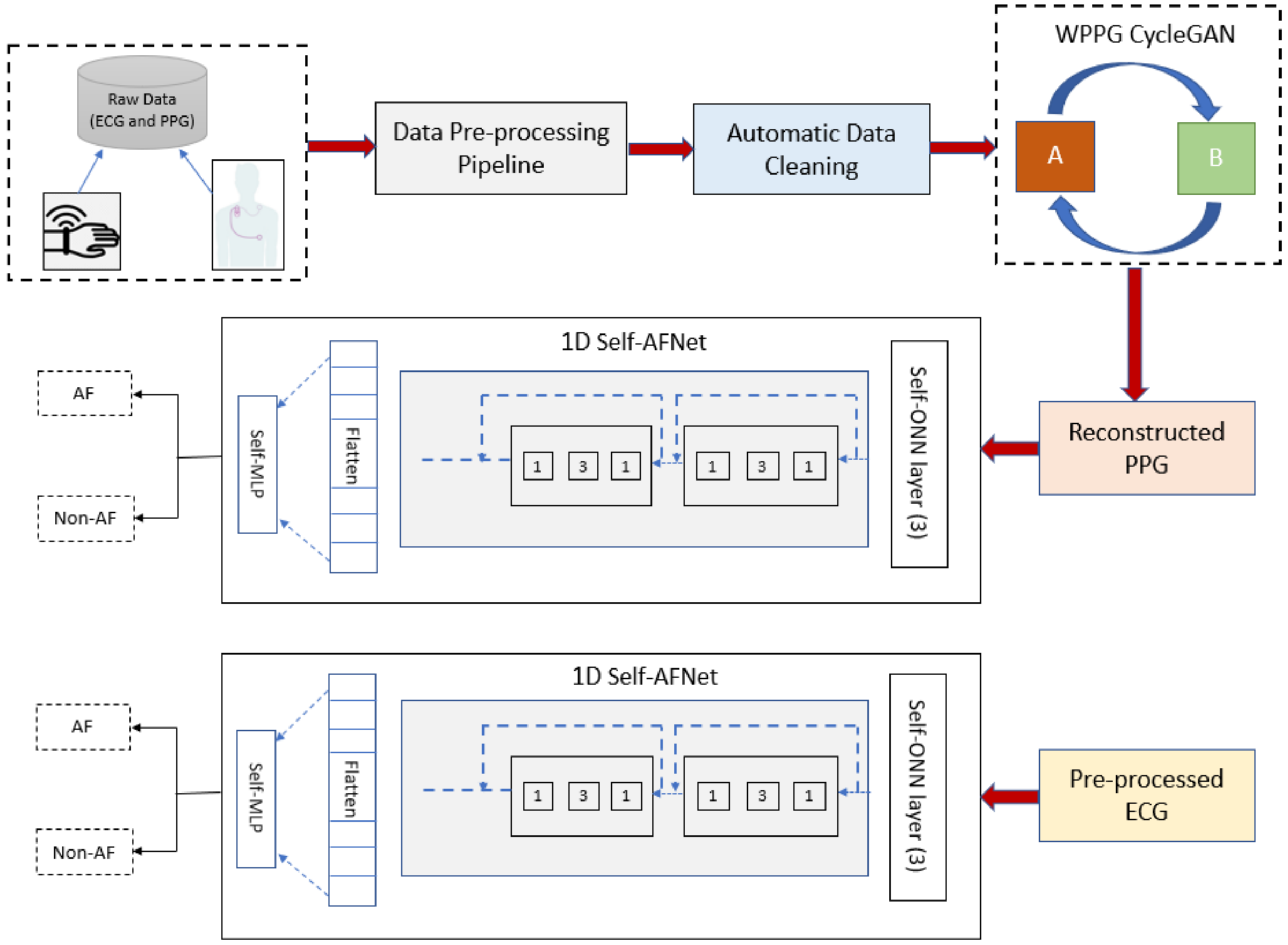

*Figure 2: Simple overview of the overall classification approach.*

### 3.1. Self-ONN

Homogenous network structure along with linear neuron model is a common drawback of CNNs [41]. This drawback is addressed in the concept of Generative Operational Perceptrons or short GOPs [41]–[44]. Operational Neural networks or ONNs incorporated the concept of GOPs [41] and later, Self-ONN, a variant of ONNs was proposed to introduce non-homogeneity in neural networks [29], [31]. Self-ONN mimics the biological neuron activity during activation or signal propagation [42], [46]. In a biological neuron, lots of neurochemical activities take place, such as non-linear synaptic connections and integration of signals in soma [41], [42]. Similarly, Self-ONN mimics this non-linear process in the neural network. "Nodal operator" or $\Psi$ in "Nodal operation" of Self-ONN is analogous to the synaptic connection of biological neurons and "Pool operator" or $P$ is analogous to the integration in the soma. Self-ONN does not need any pre-determined nodal operator set as ONN, while Self-ONN can generate any composite nodal operator which may not be a pre-defined function such as hyperbolic, sinusoids, exponentials, etc. If $x_m^n$ is the output in m$^{th}$ neuron of n$^{th}$ layer due to the input $y^{n-1}$, the calculation can be presented as

$$x_m^n = b_m^n + P\left(\sum_{i=1}^{N_{n-1}} \psi_{mi}^n\left(\omega_{mi}^n, y_i^{n-1}\right)\right) \quad (1)$$

Here, $\omega_{mi}^n$ is the weight and $b_m^n$ is the biased term. The ω is an array of parameters in $q$ dimensions made up of internal parameters and weights for each unique function. The combination of these arrays and inputs is combined using the nodal operator. Taylor Series approximation is used to estimate the nodal operator. So, the approximation is

$$f(x) = f(x_0) + \frac{f'(x_0)}{1!}(x - x_0) + \frac{f''(x_0)}{2!}(x - x_0)^2 + \cdots . + \frac{f^q(x_0)}{q!}(x - x_0)^q \quad (2)$$

$$f(x) = f(0) + \frac{f'(x_0)}{1!}(x) + \frac{f''(x_0)}{2!}(x)^2 + \cdots . + \frac{f^q(x_0)}{q!}(x)^q \quad (3)$$

$$f(x) = \omega_0 + \omega_1(x) + \omega_2(x)^2 + \cdots . + \omega_q(x)^q \quad (4)$$

In Eq. (2) and (3), first, second, and q$^{th}$ order derivative are defined as $f'$, $f''$, and $f^q$ respectively. The $x_0$ in Eq. (2) can be approximated to 0 by bounding the input between [-1,1] using Tanh activation in the previous layer. So, the weight ω can be expanded to q$^{th}$ order and $\omega_0$ is the bias coefficient, which is compensated by each neuron's bias term available in Eq. (1). Figure *3* illustrates the Self-ONN operation on an input vector with the nodal operation and pooling operation with $q$ order of *3*.

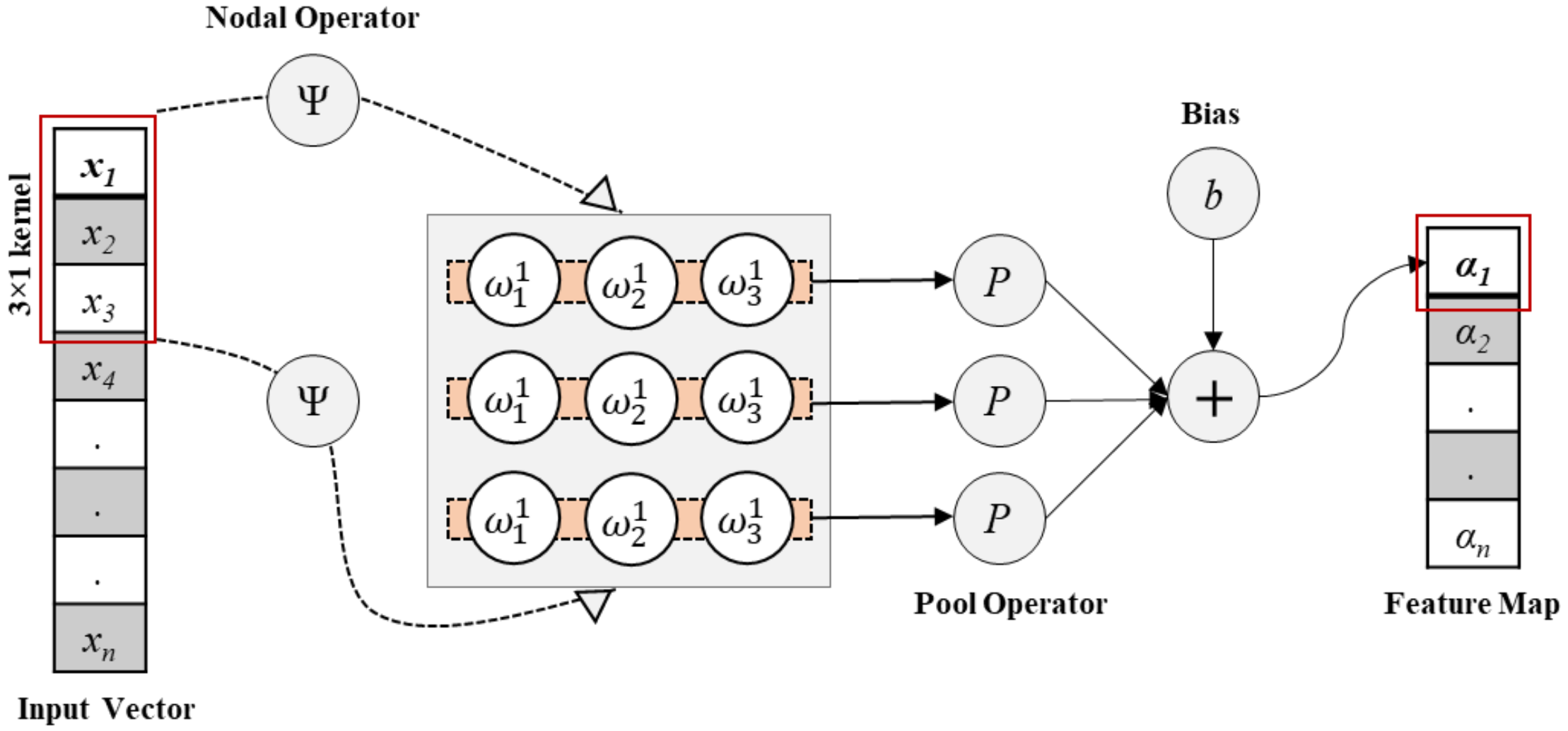

*Figure 3: Self-ONN operation on an input vector with pooling operator "P" and nodal operator "Ψ".*

### *3.2. Self-AFNet*

The architecture of state-of-the-art CNN models, such as ResNet was redesigned with Self-ONN layers in literature [45] for image super-resolution. In this research, 1D Self-ONN model [47] blocks were adopted. Figure 3 illustrates a Self-AFNet block used in this research, which is completely designed using Self-ONN layers. From Figure 4, it can be seen that pointwise and depth-wise 1D operations are employed in this Self-AFNet block. First pointwise operation was designed to increase the number of channel and second pointwise operation was designed from decreasing the number of channels to create bottleneck using kernel size of 1. So, the depth wise operation between two pointwise operations was done using 1D Self-ONN layer with kernel size of 3. The output feature from a Self-AFNet was obtained by applying a Tanh activation on the features formulated from the residual connection between input feature and bottleneck feature.

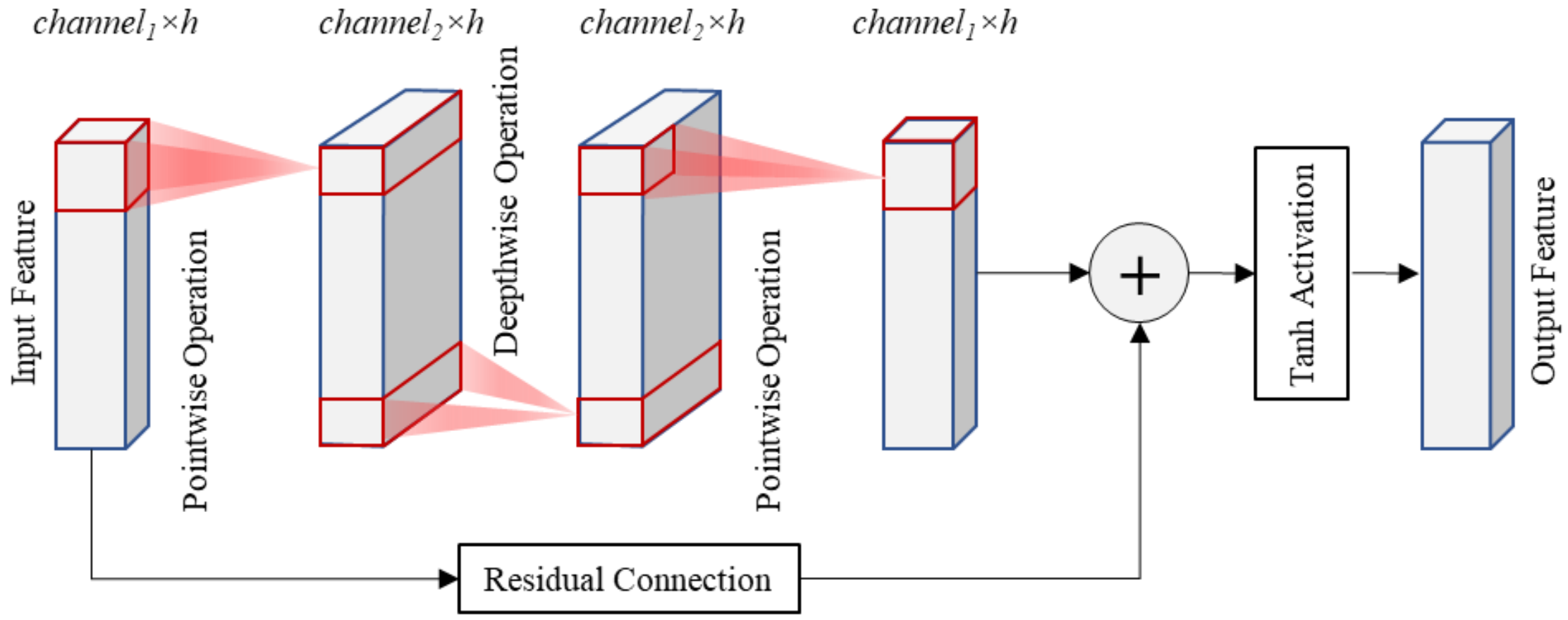

*Figure 4: Self-AFNet block designed with Self-ONN layers.*

Three Self-AFNet blocks were used to build the Self-AFNet architecture and an illustration of Self-AFNet architecture. From Figure 5, it is evident that every Self-ONN layer is followed by a BatchNorm and Tanh activation layer. The output feature after three Self-AFNet blocks is pooled to a feature size of 12 in each channel using Adaptive Average pooling. Instead of conventional multi-layer perceptron (MLP) and Self-MLP, a Self-ONN based MLP was used in this study for the classifier part. Overall, a lightweight and shallow 1D Self-ONN based Self-AFNet is proposed in this study to classify AF and Non-AF signals from wrist PPG signals.

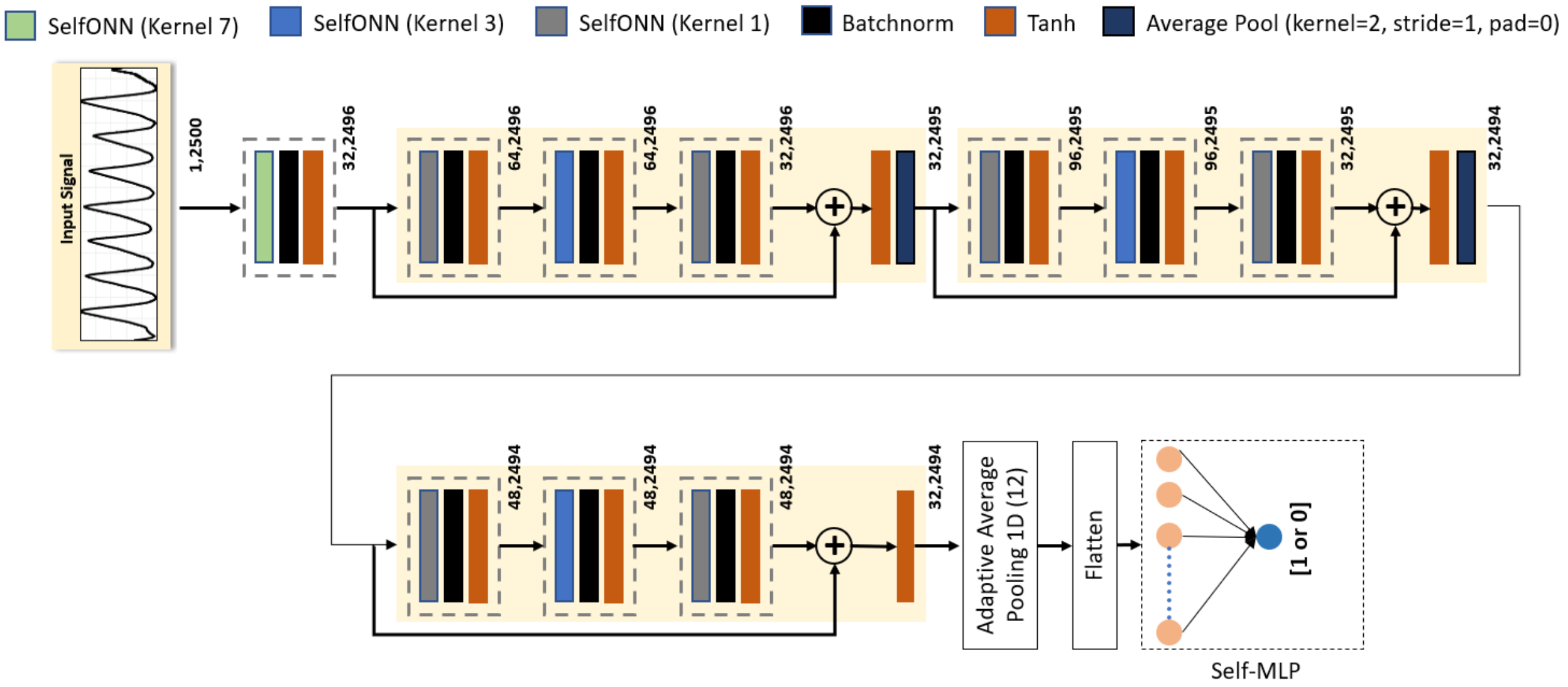

Figure 5: Self-AFNet architecture proposed in this study for classifying AF and non-AF.

### 3.3. 1D CycleGAN

Our model employs a Self-ONN based model as the discriminator and a CNN-based model as the generator. Here, we used 12 residual blocks in the generator model since it reconstructs the low-quality PPG signals better than the 9 residual block-based model. We employed an equal amount of wrist PPG signals of high and bad quality to train the operational CycleGAN for improving wrist PPG signal quality with no overlapping. We have used total of 1823 non-AF PPG signals and 707 AF PPG signals. So, a total number of 2530 noisy PPG segments and 2530 clean PPG segments are used for training the 1D-CycleGAN. The PPG segments were split into 4 folds for training and validation. We have used 70% of the PPG segments for training purposes in each fold.

The overall block diagram of our proposed CycleGAN approach is shown in Figure 6. Traditional GANs only employ adversarial loss. CycleGAN works to reduce identity and cycle-consistency loss. For the restoration of low-quality PPG signals, there are two kinds of generators used in CycleGAN. The generator GX2C learns how to reconstruct clean wrist PPG signals from noisy PPG signals that suffers from baseline wander, signal cuts, and severe motion artifacts, whereas the generator GC2X learns to produce low-quality wrist PPG signals from clean PPG signals.

In contrast, DC and DX are the analogous discriminators for GX2C and GC2X and they work to improve the adversarial loss function and direct the generators to create more accurate transformations. The adversarial loss functions are formulated in Eqn. (5) and Eqn. (6).

$$Loss_{adv1}(GX2C, DC, WPPG_X) = \frac{1}{n}\sum_{i=1}^{n}\left(1 - DC\left(GX2C\left(WPPG_X(i)\right)\right)\right)^2 \quad (5)$$

$$Loss_{adv2}(GC2X, DX, WPPG_C) = \frac{1}{n}\sum_{i=1}^{n}\left(1 - DX\left(GC2X\left(WPPG_C(i)\right)\right)\right)^2 \quad (6)$$

Here, $WPPG_X$ and $WPPG_C$ are noisy and clean wrist PPG data, respectively, and Here, 'n' is the number of segments used in the training set. The discriminator DC tries to discriminate between real clean and fake (generated) clean PPG segments and guides the generator GX2C to generate more realistic clean PPG waveforms while DX tries to do just the opposite.

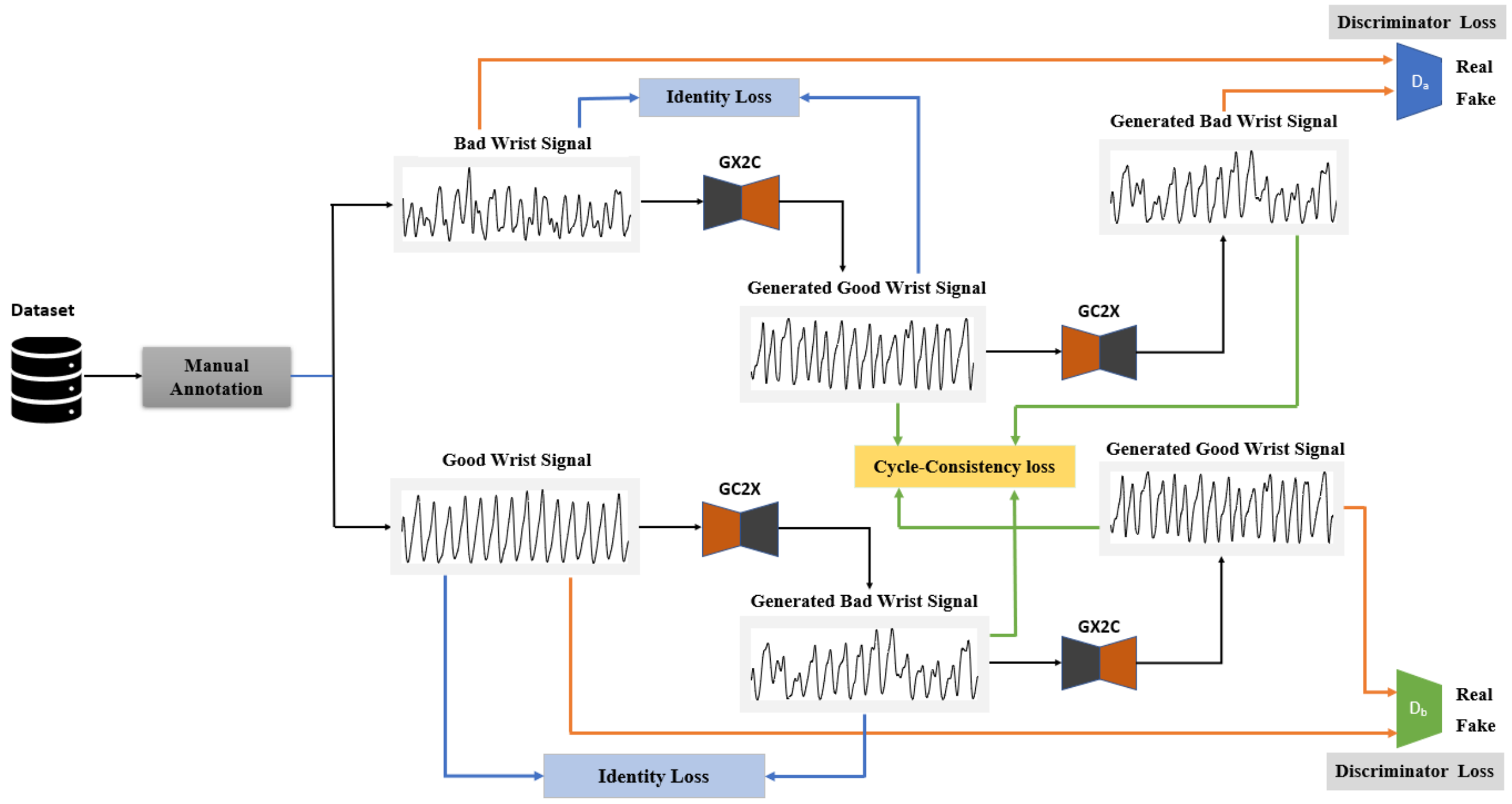

*Figure 6: Complete overview of the WPPG CycleGAN for wrist PPG signal reconstruction.*

The cycle consistency loss measures the difference between the PPG segment input to $Generator\ 1$ and the PPG segment produced by $Generator\ 2$ and the generator models are updated accordingly to reduce the difference in the PPG segments. The cycle-consistency loss combines the consistency losses from the two pathways in an effort to get more realistic end results.

$$Loss_{cyc}(GX2C, GC2X, WPPG_X, WPPG_C) = \frac{1}{n}\sum_{i=1}^{n}\left(GC2X\left(GX2C\big(WPPG_X(i)\big)\right) - WPPG_X(i)\right) + \frac{1}{n}\sum_{i=1}^{n}\left(GX2C\left(GC2X\big(WPPG_C(i)\big)\right) - WPPG_C(i)\right) \quad (7)$$

Identity loss retains the shape of the signal and prevents morphological changes in the input PPG segments when the input PPG segment is clean.

$$Loss_{ide}(GX2C, GC2X, WPPG_X, WPPG_C) = \frac{1}{n}\sum_{i=1}^{n}\left(\left(GX2C\big(WPPG_C(i)\big)\right) - WPPG_C(i)\right) + \frac{1}{n}\sum_{i=1}^{n}\left(\left(GC2X\big(WPPG_X(i)\big)\right) - WPPG_X(i)\right) \quad (8)$$

The objective of training CycleGAN is to reduce the total loss formulated in Eqn. (9)

$$Loss_{total} = Loss_{adv1} + Loss_{adv2} + \lambda Loss_{cyc} + \beta Loss_{ide} \quad (9)$$

Here, $\lambda$ and $\beta$ are loss weights that are tuned before training. The architecture of both the generator (Resnet-12 Block) and discriminator (Self-ONN Model) are shown in Figure 7.

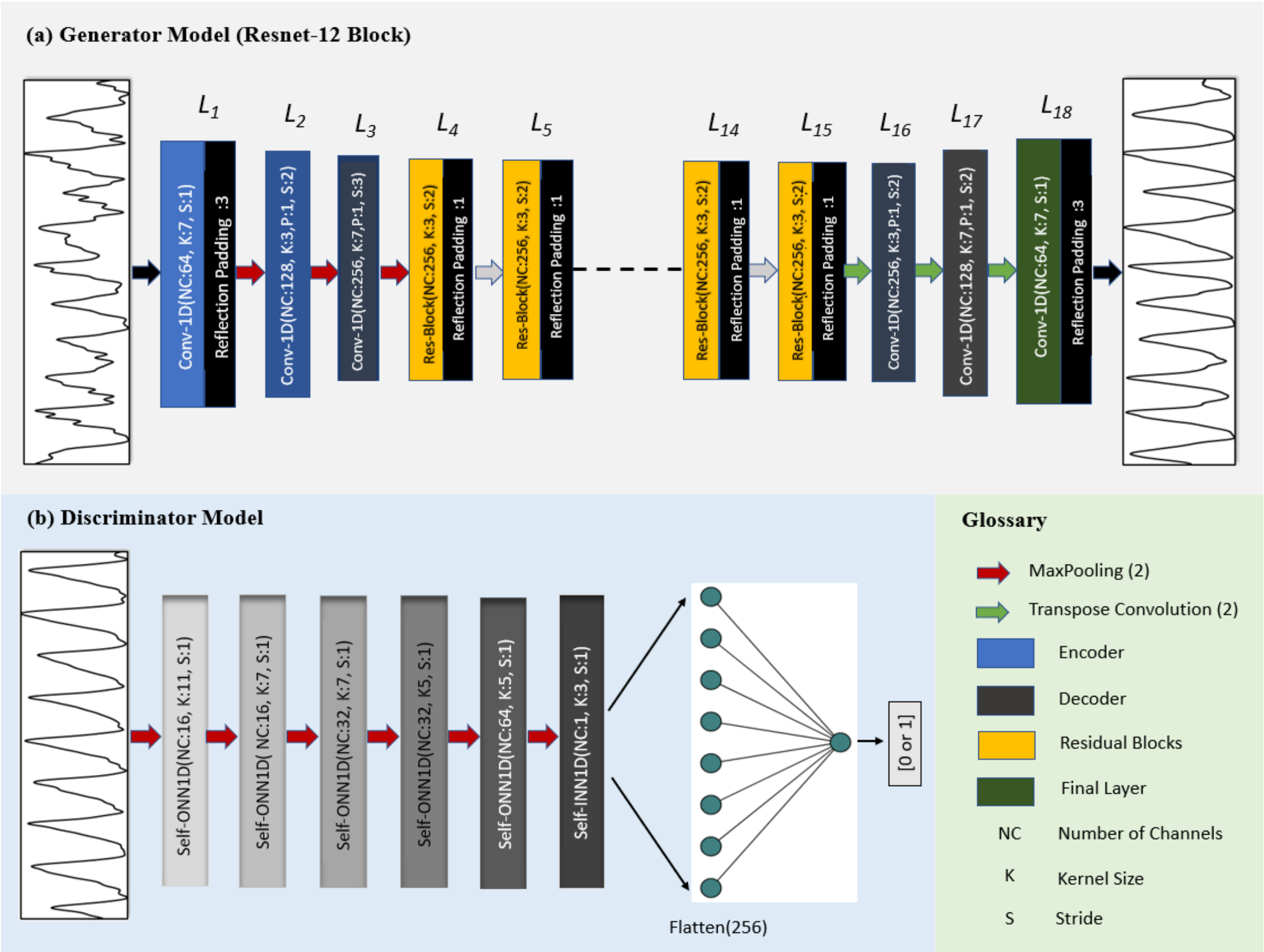

*Figure 7: Block diagram of (a) Generator (top) and (b) Discriminator (bottom left) for the proposed 1D CycleGAN.*

## *3.4. Experimental Setup*

We have implemented the 1D Self-AFNet model for classification tasks using Self-ONN layers. For classification experiments, we employed a training approach with a batch size of 16 and a number of epochs of 20. With SGD optimizer, we used a learning rate of 0.25 and different Macclaurin approximations, q=1,3,5 and 7. For CycleGAN, we have employed Resnet-12 Block as the generator and 6 layers Self-ONN model as the discriminator. For CycleGAN, we have used a training paradigm with training epoch 150 and a batch size of 16. With a lambda constant ($\lambda$) value of 0.5, we have chosen adam optimizer with a learning rate of 0.0005. Both our proposed classification model and CycleGAN were implemented in PyTorch. The FastONN library, a quick GPU-enabled toolkit created in Python and PyTorch to design and train operational neural networks, is used to implement the suggested Self-ONN layers [48].

### ***Quantitative Evaluation Metrics for Blind Restoration of Wrist PPG Signals***

We do not have any exactly clean PPG signals for corresponding low-quality signals for the quantitative evaluation of wrist PPG reconstruction using CycleGAN. As a result, we cannot use any suitable metrics such as root mean square (RMSE) or mean average error (MAE), because there is no ground truth. To measure the amount of irregularity and unpredictability of wrist PPG signals, we have provided several entropy functions such as Fuzzy Entropy [49], Sample Entropy (SampEn) [50], and Approximate Entropy (ApEn) [51]. It is worth mentioning here that entropy has no negative values, while a positive value refers to disorder in the signal of interest.

Any periodic system has an entropy of zero, and it remains invariant when scaled or translated [51]. Pincus et al. [51] have proposed ApEn to quantify the amount of irregularity in different time series data and physiological signals (e.g. ECG and Electromyography (EMG)). To compute irregularity and distinguish healthy signals from abnormal ones, ApEn can be defined as Eq. (10)

$$ApEn = \phi^m(r) - \phi^{m+1}(r) \tag{10}$$

Now, $\phi^m(r)$ can be formulated as in Eq. (11),

$$\phi^m(r) = (N - m + 1)^{-1} \sum_{i=1}^{N-m+1} ln \cdot C_i^m(r) \tag{11}$$

Within a tolerance r, $C_i^m(r)$ measures the irregularity. $\phi^m(r) - \phi^{m+1}(r)$ iterates through the data points, measuring the mean stability of those patterns. Richman et al. [50] suggested SampEn as an alternative to ApEn which is defined in Eq. (12),

$$SampEn(m, n, r) = -\ln\left(\frac{A^m(r)}{B^m(r)}\right) \tag{12}$$

Where, B is defined as the overall number of templates matches with length m, and A is defined as the overall number of forward matches with length m+1.

Here, $A^m(r)$ and $B^m(r)$ are defined in Eqs. (13) and (14), respectively.

$$A^m(r) = (N - m)^{-1} \sum_{i=1}^{N-m} A_i^m(r) \tag{13}$$

$$B^m(r) = (N - m)^{-1} \sum_{i=1}^{N-m} B_i^m(r) \tag{14}$$

In this scenario, $A_i^m(r)$ and $B_i^m(r)$ quantifies the level of irregularity within a certain tolerance r for a signal with N data points, while m represents the length of the compared data runs (i.e., the threshold for checking patterns). $r$ can alternatively be defined as the filter level or the threshold level for detecting anomalies (i.e., any irregularity or abnormality with an amplitude lower than this threshold will be ignored). A new measure of similarity or regularity, Fuzzy Entropy (FuzzyEn) was introduced by Chen et al. [49] for measuring time series irregularities and was used to describe surface EMG (sEMG) signals. Fuzzy Entropy is defined in Eq. (15)

$$FuzzyEn = \ln\left(\phi^m(n, r)\right) - \ln\left(\phi^{m+1}(n, r)\right) \tag{15}$$

$\phi^m$ in the FuzzyEn can be defined as Eq. (16),

$$\phi^m(n, r) = (N - m)^{-1} \sum_{i=1}^{N-m} \left( (N - m - 1)^{-1} \cdot \sum_{j=1, j\neq 1}^{N-m} D_{ij}^m \right) \tag{16}$$

Here, $D_{ij}^m$ denotes the degree of similarity or regularity between two adjacent vectors $X_i^m$ and $X_j^m$.

In this study, we have used N = 2500 data samples, m = 2, and r = 0.1 for all entropy measuring functions, which are conventional values from the literature [49]–[53]. According to the literature, m should be chosen so that N is between $10^m$ and $30^m$ for reliable quality assessment of irregularity. This requirement was met based on the numbers we chose.

We also used Permutation Entropy (PermEn), which may be used indiscriminately in real-world applications due to its fast speed and robustness, making it the method of choice for applications involving massive databases.

PermEn can be computed using the following Eq. (17)

$$H(n) = -\Sigma P(\pi) \log(\pi) \tag{17}$$

Where P(π) represents the relative frequency and can be formulated as in Eq. (18)

$$p(\pi) = \frac{\#\{t | t \leq T - n, (x_{t+1}, \dots, x_{t+n}) \text{has type } \pi\}}{T - n + 1} \tag{18}$$

Initially, preprocessed raw low-quality wPPG signals were sent to the generator of the CycleGAN, and we get the restored PPG signals (Pass 1). Then, the restored wPPG signals are passed through the trained generator again (Pass 2) to see whether 2nd pass improves signal quality or not.

### *Performance Evaluation for Classification*

Here, we will present the evaluation metrics used in our AF classification study using Self-AFNet model. We utilized accuracy, sensitivity, precision, recall and F1 score, receiver operating characteristic (ROC) curves and area under the curve (AUC) value. AUC value ranges from 0 to 1. The higher the AUC, the better the performance of the model at distinguishing between the two classes. The mathematical representation of these performance indicators is shown below:

$$Overall\ accuracy = \frac{\text{TP+FN}}{\text{TP+TN+FP+FN}} \times 100 \quad (19)$$

$$Precision = \frac{\text{TP}}{\text{TP+FP}} \times 100 \quad (20)$$

$$Recall = \frac{\text{TP}}{\text{TP+FN}} \times 100 \quad (21)$$

$$Specificity = \frac{\text{TP}}{\text{FP+FN}} \times 100 \quad (22)$$

$$F1\ Score = \frac{2 \times precision \times recall}{\text{precision+recall}} \times 100 \quad (23)$$

Here, TP = true positive, TN = true negative, FP=false positive, FN = False negative

Since, the number of AF and non-AF segment classes are not same, we have also used weighted metrics of accuracy, precision, recall, sensitivity and F1 score.

## 4. Results and Discussions

As long-term continuous labeled wrist PPG datasets for AF identification tasks are rarely made public, it was challenging to get a suitable dataset. We managed to get access to the OpenAIRE dataset [36] and used it for training and evaluation for our proposed framework. This section shows the results of two-split investigation of the dataset for robust AF detection.

### *4.1. Corrupted Signal detection*

With clean and bad quality wPPG signals, there are many invalid and corrupted wPPG signals that has no PPG like pattern and higly corrupted due to extensive motion artefacts. It takes a long time to manually remove those corrupted wPPG segments and is not practical for real-world applications as well. Therefore, we first trained a 1D Self-AFNet classifier using the limited number of PPG segments to detect corrupted PPG signals.

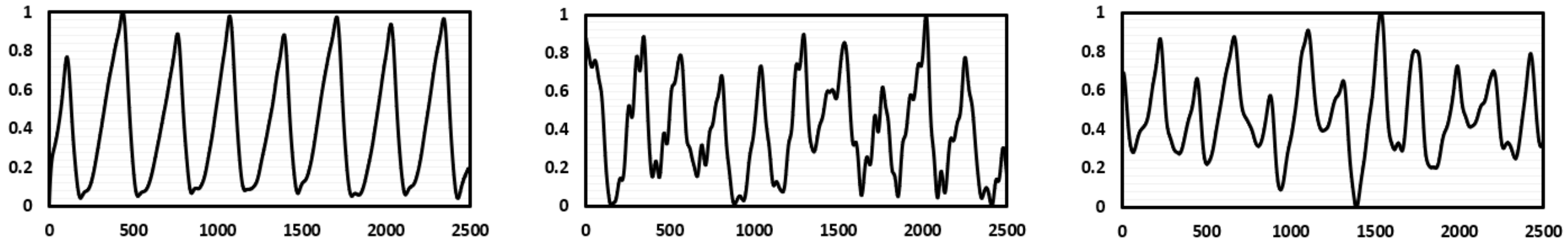

**(a)**

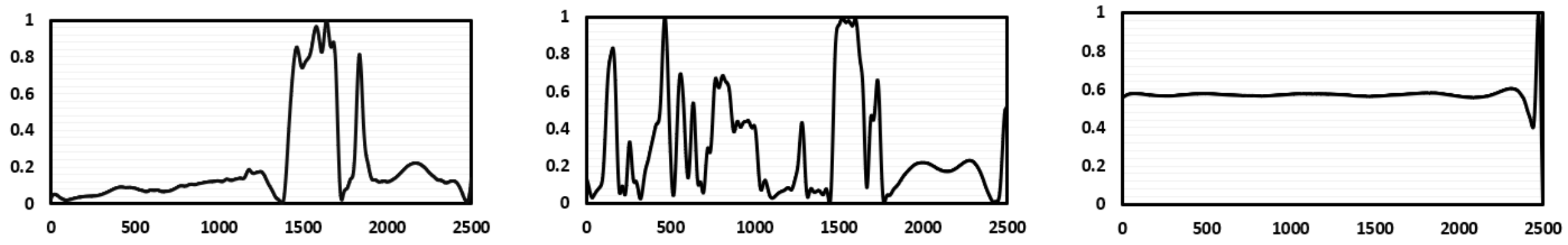

**(b)**

*Figure 8: Visualization of acceptable PPG segments (a) and corrupted PPG segments (b).*

We have found an overall accuracy of 92.09% using 13,067 PPG segments on 5-fold cross-validation method as shown in Table 2. Then, we applied this model in the remaining PPG dataset to remove the corrupted PPG signals (Figure 8). Only 21,278 wPPG signals segments were found acceptable from 40,115 wPPG segments.

**Table 2:** Performance measure of detecting invalid wrist PPG segments

| Classifier | | Accuracy | Precision | Sensitivity | F1_score | Specificity |
|---|---|---|---|---|---|---|
| Self-AFNet (q=3) | **Corrupted** | 92.09 | 93.86 | 89.09 | 91.41 | 94.78 |
| | **Acceptable** | | 90.64 | 94.78 | 92.66 | 89.09 |
| | **Weighted Average** | | 92.16 | 92.09 | 92.07 | 91.78 |

## *4.2. Blind Restoration of Wrist PPG and Quantitative Evaluation*

Quantitative evaluation of wPPG restored using 1D-CycleGAN is demonstrated with four entropy metrics as shown in Table 3. It is evident from all the metrics that the restoration process has decreased the entropies from the raw data. The most significant difference was observed for PermEn, which is suitable for a large number of signal samples.

**Table 3:** Evaluation of our proposed WPPG 1D CycleGAN

| | **FuzzyEn** | **SampEn** | **ApEn** | **PermEn** |
|---|---|---|---|---|
| **Raw Original** | 0.0170 | 0.1827 | 0.2316 | 0.9507 |
| **Restored (Pass 1)** | 0.0161 | 0.1615 | 0.2124 | 0.7424 |
| **Restored (Pass 2)** | 0.0160 | 0.1587 | 0.2169 | 0.7352 |

Qualitative results show that degree of irregularity and noise decreases using reconstructed PPG signals. During reconstruction, CycleGAN performs a variety of operations. Entropy may rise sometimes since CycleGAN produces vivid signals sometimes from extremely smooth waveforms to remove severe motion artifacts and signal cuts. We now give a visual representation of the quality of the wrist PPG signals that have been reconstructed using our proposed 1D-CycleGAN model. Figure 9 presents non-AF signals along with their corresponding reconstructed signals (after 1$^{st}$ and 2$^{nd}$ passes), as well as AF signals along with their corresponding reconstructed signals.

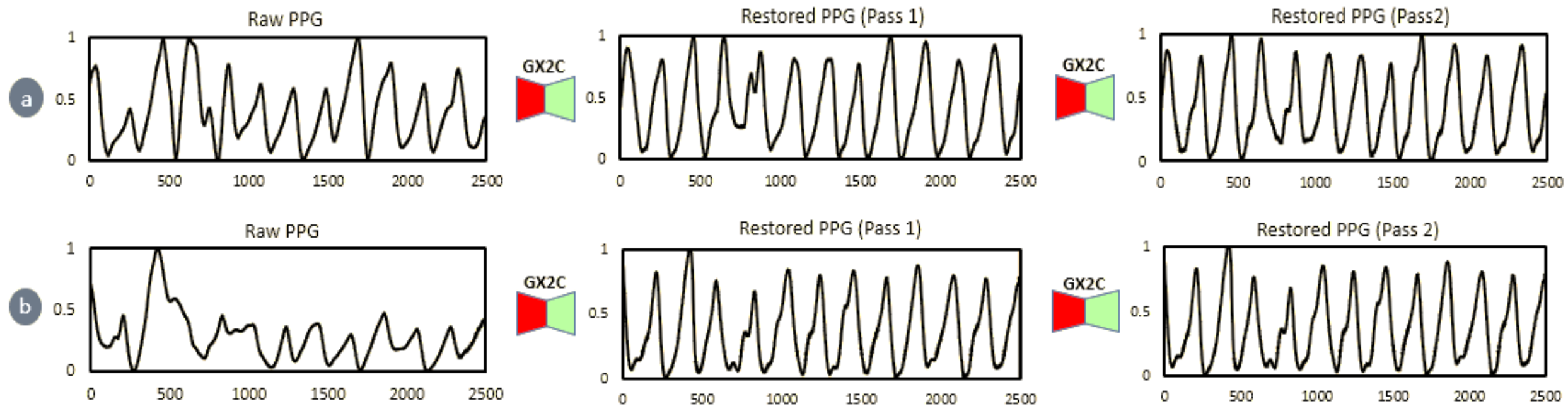

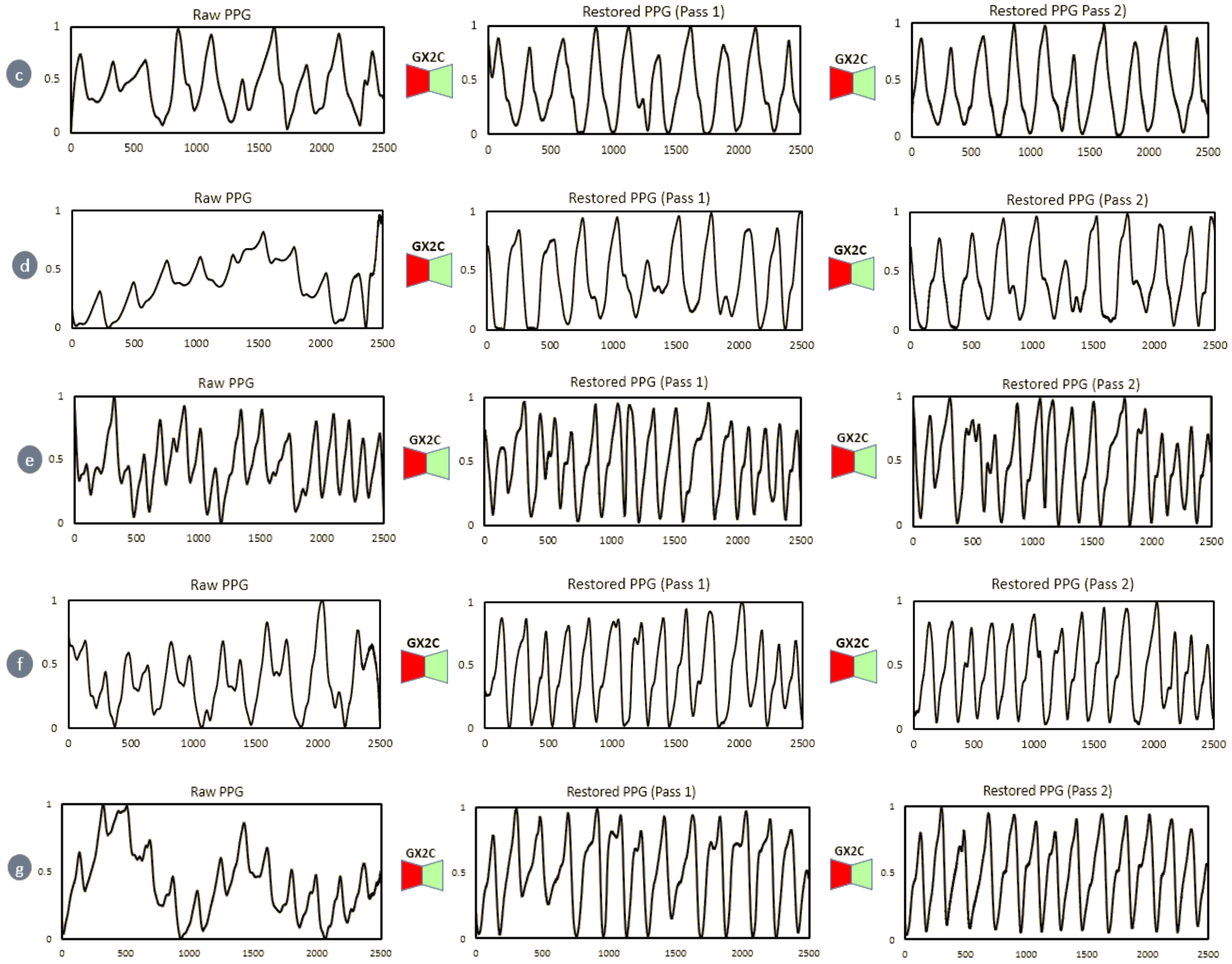

*Figure 9: Visualization of the performance of WPPG CycleGAN with raw signals, pass 1 (restored signals), and pass 2 (restored signals). Here, (a)- (d) represents the non-AF PPG signal restoration performance and (e)-(g) shows the AF signal restoration with our proposed 1D CycleGAN.*

## *4.3. Experiments and Evaluation*

We carried out two experiments to compare the 1D Self-AFNet model's performance in real-world circumstances and demonstrate the robustness of the model. The reconstructed PPG segments of Test Split 1 and 2 are tested in such a way that one split was used for training while other split was used for testing and this was reversed and the final results are reported in Table 4. We have conducted a number of experiments using the Self-AFNet model for classification tasks with various q-order for the Self-ONN layers. It should be noted that each generative neuron in an operational layer has the ability to enhance each kernel's nodal operator function. Heterogeneous diversity at the neuron level subsequently improves network diversity and classification performance. For q=1, the Self-AFNet becomes a CNN model while q=3, 5, and 7 make it operational model. From Table 4, it is clear that Self-ONN layers with q=3, 5, and 7 in the Self-AFNet model perform better than traditional CNN layers where q=1 in the Self-ONN layers.

**Table 4:** Performance comparison of the Self-AFNet model with different Q orders using reconstructed wPPG signals

| Classifier | Evaluation | Accuracy | Precision | Sensitivity | F1_score | Specificity |
|---|---|---|---|---|---|---|
| **Self-AFNet_Q1(CNN)** | Test Split 2 | 95.69 | 95.84 | 95.69 | 95.74 | 94.61 |
| **Self-AFNet_Q3** | | **97.09** | **97.08** | **97.09** | **97.06** | **92.98** |
| **Self-AFNet_Q5** | | 96.82 | 96.81 | 96.82 | 96.81 | 94.14 |
| **Self-AFNet_Q7** | | 96.94 | 96.93 | 96.94 | 96.91 | 92.63 |
| **Self-AFNet_Q1(CNN)** | Test Split 1 | 94.58 | 94.55 | 94.58 | 94.49 | 88.12 |
| **Self-AFNet_Q3** | | **96.41** | **96.39** | **96.41** | **96.4** | **93.59** |
| **Self-AFNet_Q5** | | 96.34 | 96.33 | 96.34 | 96.34 | 93.92 |
| **Self-AFNet_Q7** | | 96.31 | 96.33 | 96.31 | 96.31 | 94.44 |

We have compared the performance of Self-AFNet model with reconstructed PPG data using CycleGAN to original preprocessed PPG segments. We have found significant improvement in AF classification using reconstructed PPG signals from CycleGAN. We demonstrate the results of these two experiments with both raw original PPG signals and reconstructed PPG signals by CycleGAN in Table 5.

**Table 5:** Performance comparison between raw and reconstructed wPPG and ECG Test Splits using Self-AFNet (q=3)

| Data | Evaluation | | Accuracy | Precision | Sensitivity | F1_score | Specificity |
|---|---|---|---|---|---|---|---|
| **Raw wPPG** | Test Split 2 | Non-AF | **92.01** | 97.06 | 92.29 | 94.61 | 91.13 |
| | | AF | | 78.83 | 91.13 | 84.53 | 92.29 |
| | | Weighted Average | | **92.69** | **92.01** | **92.19** | **91.41** |
| | Test Split 1 | Non-AF | **93.47** | 94.03 | 97.41 | 95.69 | 82.02 |
| | | AF | | 91.59 | 82.02 | 86.54 | 97.41 |
| | | Weighted Average | | **93.41** | **93.47** | **93.35** | **85.96** |
| **Reconstructed wPPG** | Test Split 2 | Non-AF | **97.09** | 97.24 | 98.98 | 98.1 | 91.09 |
| | | AF | | 96.57 | 91.09 | 93.75 | 98.98 |
| | | Weighted Average | | **97.08** | **97.09** | **97.06** | **92.98** |
| | Test Split 1 | Non-AF | **96.41** | 97.3 | 97.89 | 97.59 | 92.11 |
| | | AF | | 93.75 | 92.11 | 92.92 | 97.89 |
| | | Weighted Average | | **96.39** | **96.41** | **96.4** | **93.59** |
| **ECG** | Test Split 2 | Non-AF | **98.97** | 98.97 | 99.55 | 98.74 | 99.14 |
| | | AF | | 98.97 | 98.1 | 99.31 | 98.7 |
| | | Weighted Average | | **98.97** | **98.98** | **98.97** | **98.97** |
| | Test Split 1 | Non-AF | **98.07** | 97.9 | 99.55 | 98.72 | 93.78 |
| | | AF | | 98.62 | 93.78 | 96.14 | 99.55 |
| | | Weighted Average | | **98.08** | **98.07** | **98.06** | **95.26** |

We have found 97.09% accuracy with ~5% improvement utilizing reconstructed PPG data compared to original preprocessed signals when using Test Data 2 as unseen data (Table 5). When we used Test data 1 as unseen data, we found that AF classification performance also increases by ~3%. Here we can see a significant improvement (10.09%) in sensitivity using reconstructed AF signals compared to raw original PPG signals. Considering both experiments using Test Split 1 and 2, the overall classification performance of AF detection significantly improves on reconstructed PPG signals. Figure 10 shows confusion matrices for raw and restored wPPG signals using Test Split 1 and 2. From Figure 10 (a) and (c), it is clear that using the restored version of wPPG, our proposed model managed to predict additional 584 Non-AF signals correctly for Test Split 2. It is worth noting that this restoration did not degrade the performance of the AF detection as only one sample was additioanlly miss-classified by the model.

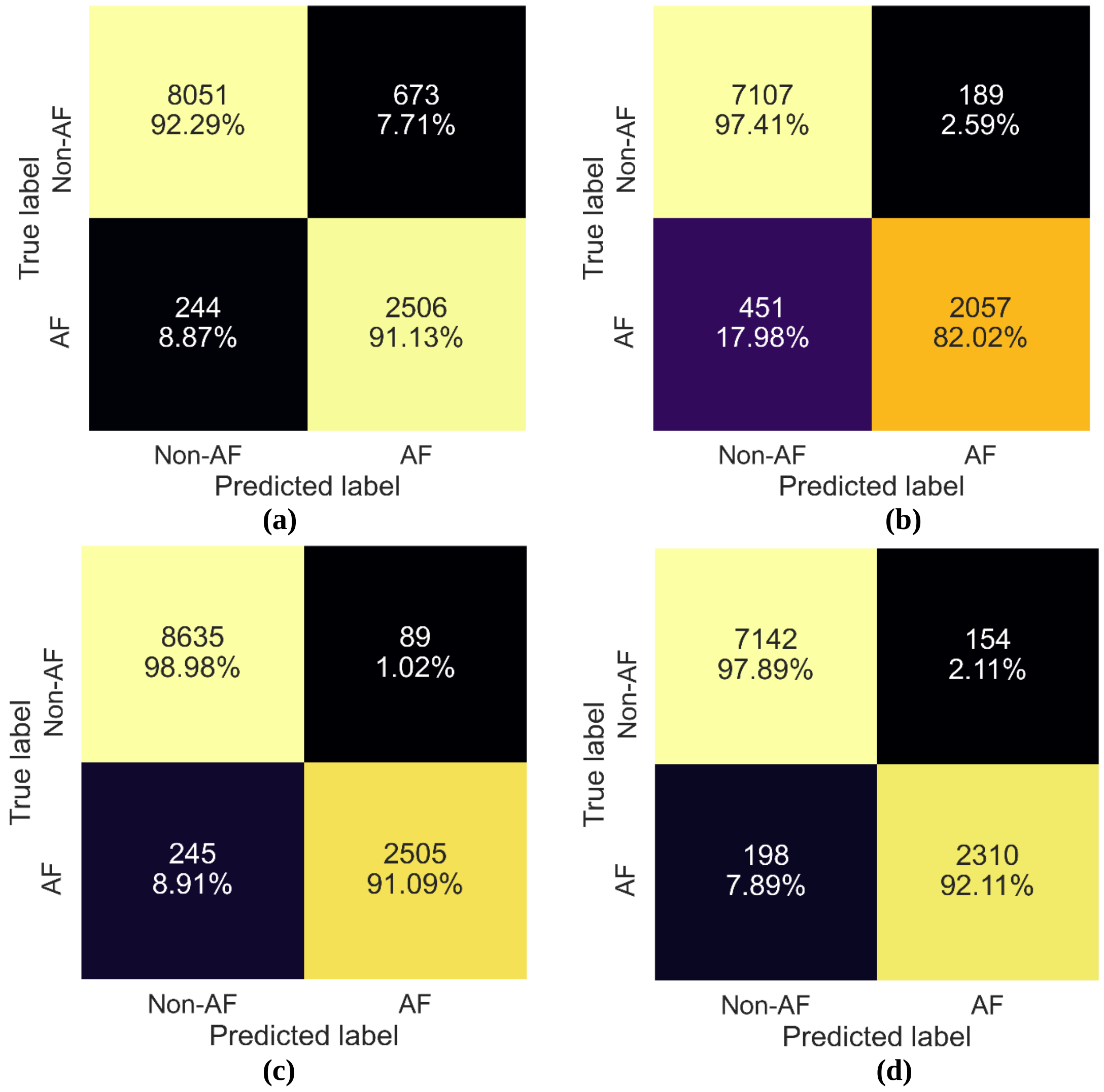

*Figure 10: Confusion matrix for raw PPG test split 2 (a), and test split 1 (b); restored PPG test split 2 (c) and test split 1 (d).*

Similarly, from Figure 10 (b) and (d), it is apparent that for the Non-AF signals classification, after restoration additional 35 samples our model can correctly predict while for the AF signal classification, after restroration additional 253 samples were correctly predicted by the model for Test Split 1, which is a boost of ~10% in sensitivity for the AF class detection.

For the AF identification task, we have also used ECG data that corresponded to wrist PPG signals. Similar to wrist PPG signals, we have divided the ECG signals into two test segments for evaluation. ECG signals are more reliable for any task involving the detection of abnormal cardiac rhythms because they are relatively clear and less noisy than wrist PPG signals. We have carried out an ECG-based AF classification task to test the reliability of our PPG-based AF classification pipeline. We present the performance of AF classification using ECG splits in Table 4. We have achieved the best performance using the Self-AFNet model with q=3 value for the operational layers. It is evident that restored wPPG utilizing CycleGAN for AF classification closely matches the ECG-based AF classification performance.

We have achieved an excellent AUC value of 0.999 using q=3 in Self-AFNet on ECG segments. Our proposed method can reliably differentiate non-AF and AF patients using ECG signals (Figure 11 and Figure 12). On the contrary, we have obtained excellent AUC values of 0.987 and 0.984 for the reconstructed test PPG splits 1 and 2, respectively. It is a strong indication that reconstructed PPG segments generated by our proposed 1D CycleGAN can be used to successfully distinguish between AF and non-AF samples.

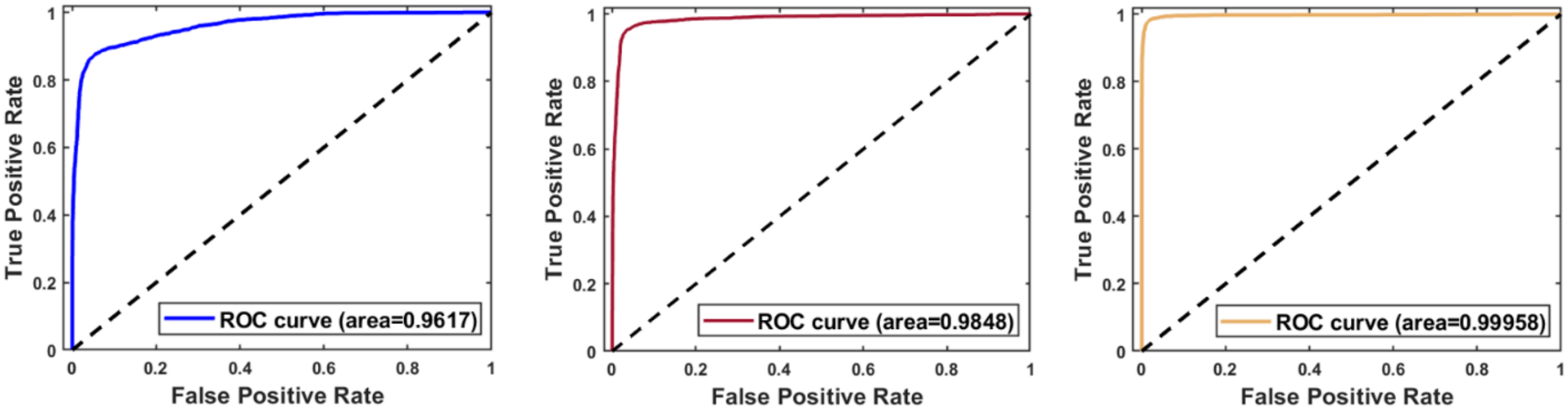

**(a)** **(b)** **(c)**

*Figure 11: ROC curves utilizing PPG signals of raw test segment 1 (a) and restored test segment 1 (b); ECG signals of test segment 1 (c)*

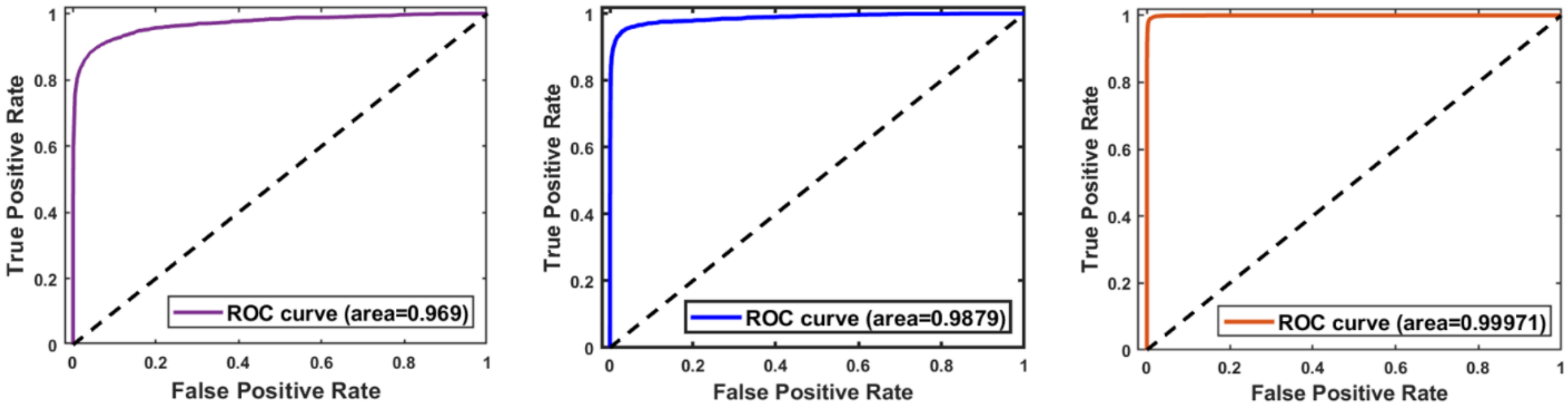

**(a)** **(b)** **(c)**

*Figure 12: ROC curves utilizing PPG signals of raw test segment 2 (a) and restored test segment 2 (b); ECG signals of test segment 2 (c)*

Finally, we provide a comparative analysis of the classification performance of wrist PPG and ECG segments. We can see that our robust AF classification performance pipeline employing CycleGAN closely matches with gold-standard ECG signals as shown in Figure 13.

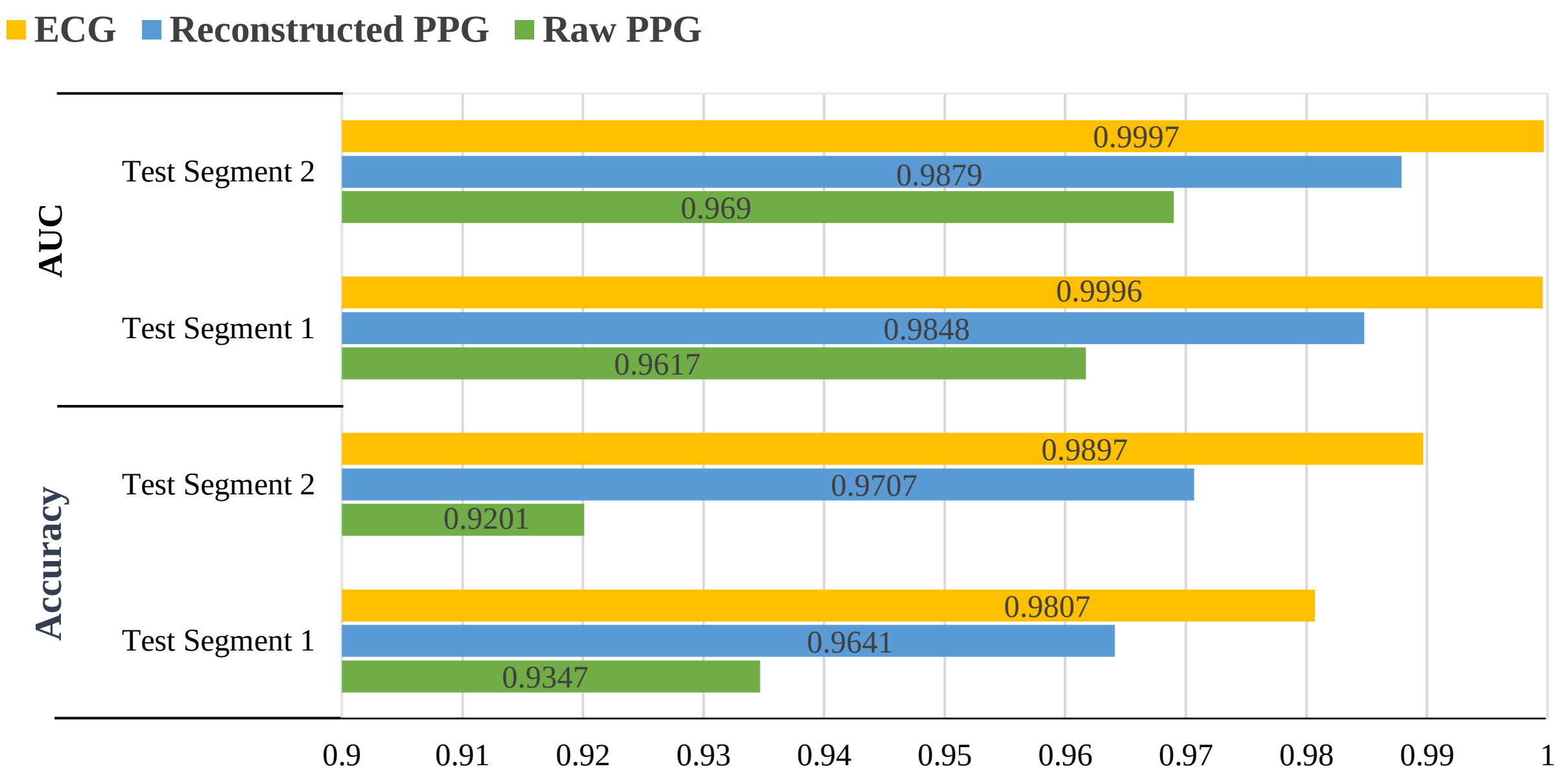

*Figure 13: Performance comparison of wrist PPG, restored wrist PPG, and ECG with AF and non-AF signals.*

Table 6 shows some recent studies of AF classification approaches using PPG data collected from wrist-worn devices and pulse oximeter devices. In our work, a good collection of wrist PPG segments for AF identification was used, comprising both clean and significantly noisy PPG signals. The proposed method[1] refers to training on test split 2 and evaluation on test split 1. On the other hand, the proposed method[2] indicates training on test split 1 and test evaluation on test split 2. Only Nguyen et al [54] and Cheng et al. [55] slightly outperform our model, however Nguyen et al [54] used only clean wPPG signals while Cheng et al. [55] has used both clean wrist and finger PPG signals. Therefore, the comparison of these works are not fair our work as we have used clean and noisy wPPG signals which are very challenging to classify reliably. Thus, it is evident that our restoration and classification framework using wrist PPG signals has achieved the most robust performance.

**Table 6:** Comparison with the state-of-the-art works in the literature of PPG-based AF classification.

| Authors | Segment Length | PPG Type | Segments | Approach | Evaluation matrix |
|---|---|---|---|---|---|
| Fahimeh et al. [27] | 30s | Wrist+Finger | 12,395 | Ensemble based Feature Selection and ML model | ACC 93%, Sen 89%, and Spe 97% |
| Bashar et al. [26] | 30s | Wrist | 366 (clean) | ML model with MNA detection, PAC/PVC detection | ACC 97.54%, Sen 98.13%, and Spe 97.43% |
| Nguyen et al [54] | 30s | Wrist | 2,542 (clean) | CNN Models with LSTM and RRI,PPI | ACC 98.08%, Sen 96.82% and Spe 98.86% |
| Ramesh et al. [19] | 30s | Wrist | 246 (clean) | 1D CNN with HRV-derived features | ACC 95.10%, Sen 94.60% and Spe 95.20% |
| Cheng et al. [55] | 10s | Wrist+Finger | 95.5h (total) | Time-frequency chromatograph with CNN-LSTM model | ACC 98.21%, Sen 98.0%, and Spe 98.07% |
| **Proposed Method[1]** | 10s | Wrist | **21,278** | 1D Self-AFNet with 1D CycleGAN Model | **ACC 96.41%** |
| **Proposed Method[2]** | | | | | **ACC 97.09%** |

**ACC refers to accuracy*

## 5. Conclusion

This study presented a novel approach for classifying atrial fibrillation without using any patient-specific information. We have employed the state-of-the-art pre-processing techniques for automatic acceptable signal detection and AF signal restoration and proposed a robust 1D Self-AFNet model for AF classification. ECG acts as a gold standard for the reliable atrial fibrillation detection task. Since real-time long-term monitoring of ECG signals is difficult or nonconformable to the users, wrist PPG signals non-invasively collected by smartwatches can be extensively used to detect AF. Two experiments were carried out to provide a comparative analysis of robust AF detection in real-world applications. In addition, a novel 1D Self-AFNet model was proposed after conducting a series of experiments with different Maclaurin approximations (i.e., different q values (1, 3, 5, and 7)). The best performance on both ECG and wrist PPG segments was obtained using the order of q=3 for the Self-AFNet model. This is the first study attempting to remove the wrist PPG signal noises blindly and restore wrist PPG signals with the aid of an operational 1D CycleGAN. Results demonstrate that the performance of the proposed AF classification pipeline significantly improves using reconstructed PPG signals and it closely matches the AF detection performance using ECG signals. Though the current study is focused on AF detection from wearable devices, PPG signals may also be used to measure blood oxygen levels, systolic blood pressure (BP), cardiac output, and respiratory rate. Poor signal quality due to motion artifacts when users are moving their arms or walking or running, the AF detection with Wrist PPG is greatly hampered if only conventional classification techniques are used, no matter how robust the classification algorithm is, whereas a restored version of wrist PPG signal can be used to detect AF reliably using comparatively shallow Self-AFNet model (46,418 trainable parameters). We can use the 1D version of CycleGAN for the reconstruction of PPG signals to eliminate artifacts from the wrist PPG signals. In future work, we can create a robust pipeline to extract morphological features for measuring respiratory rate [56], [57], and blood oxygen levels [58] and can also be used to efficiently extract information about disorders of the cardiovascular system [59] and detect several other cardiac arrhythmias [60].

**Funding**

This research is supported via funding from Prince Sattam Bin Abdulaziz University project number (PSAU/2023/R/1445). The statements made herein are solely the responsibility of the authors. The open access publication cost is covered by Qatar National Library.

**Data availability statement**

The compiled dataset used in this study can be made available upon a reasonable request to the corresponding author.

**Conflicts of Interest**

Authors do not have any conflict of interest to declare.